%% file: coverageHole.tex
\documentclass[conference,10pt]{IEEEtran}
\usepackage[ruled,vlined,boxed]{algorithm2e}
\usepackage{tikz}  
\usetikzlibrary{shapes,arrows}
\usepackage{amsmath}
\usepackage{amsthm}
\usepackage{blindtext}
\usepackage{algpseudocode}
\usepackage{algorithm2e}
\usepackage{algcompatible}
\providecommand{\keywords}[1]
{
  \small	
  \textbf{\textit{Keywords---}} #1
}
\usepackage[english]{babel}
\usepackage{siunitx}
\usepackage{graphicx}
\renewcommand{\figurename}{Figure.}
\usepackage{hyperref}
\usepackage{float}
\usepackage[nolist]{acronym}
\usepackage{comment}

\ifCLASSOPTIONcompsoc
 \usepackage[caption=false,font=normalsize,labelfont=sf,textfont=sf]{subfig}
\else
 \usepackage[caption=false,font=footnotesize]{subfig}
\fi
\begin{document}

\title{ Coverage Hole Elimination System  in Industrial Environment}
\author{\IEEEauthorblockN{Mervat Zarour\IEEEauthorrefmark{1}, Shreya Tayade\IEEEauthorrefmark{1}, Sergiy Melnyk\IEEEauthorrefmark{1} and Hans D. Schotten\IEEEauthorrefmark{1}\IEEEauthorrefmark{2}}
\IEEEauthorblockA{\IEEEauthorrefmark{1}Intelligent Networks Research Group,
German Research Center for Artificial Intelligence, Kaiserslautern, Germany\\$\{$mervat.zarour, shreya.tayade, sergiy.melnyk, Hans$\_$Dieter.Schotten$\}$@dfki.de}
\IEEEauthorrefmark{2}Department of Wireless Communication and Navigation, Technical University of Kaiserslautern, Kaiserslautern, Germany\\schotten@eit.uni-kl.de}

\maketitle
 
\input{IEEEtran/acroList}
\begin{abstract}

The paper proposes a framework to identify and avoid the  coverage hole in an indoor industry environment. We assume an edge cloud co-located controller that followers the \ac{agv} movement on a factory floor over a wireless channel. The coverage holes are caused due to blockage, path-loss, and fading effects. An \ac{agv} in the  coverage hole may lose connectivity to the edge-cloud and become unstable. To avoid connectivity loss, we proposed a framework that identifies the position of coverage hole using a \ac{svm} classifier model and constructs a binary coverage hole map incorporating the \ac{agv} trajectory re-planning to avoid the identified coverage hole. The \ac{agv}'s re-planned trajectory is optimized and selected to avoid coverage hole the shortest coverage-hole-free trajectory. We further investigated the look-ahead time's impact on the \ac{agv}'s re-planned trajectory performance. The results reveal that an \ac{agv}'s re-planned trajectory can be shorter and further optimized if the coverage hole position is known ahead of time.
\end{abstract}
\IEEEpeerreviewmaketitle
\section{Introduction}  
In Industry 4.0,  edge-cloud computing technology in wireless networked control systems receives growing attention. The major advantage in this context is the ability to collect and process the massive amount of data generated by sensors and actuators at the network's edge. This ensures real-time data processing and facilitates the controller to react to industrial issues. The real-time requirement in the edge cloud control system depends on the quality of the wireless connection~\cite{cs_req}. So the high-reliability performance of the wireless link in an industrial environment is essential as the signal may be blocked due to many large machines and metal surfaces. Also, the environmental structure on the industry floor leads to poor signal propagation due to shadowing, blocking, and reflection effects. The poor received signal power has the potential to cause an interruption in the radio channel and coverage holes in the radio power map. Nevertheless, shadow fading and multi-path fading behave as stochastic responses depending on the dynamics of the environment  depending  on the obstacle position or environment layout change. The radio map for the whole propagation environment is then required. The driving test is not just time intensive, but it is computationally expensive ~\cite{TD}. The 3rd Generation Partnership Project (3GPP) proposed a minimized test drive using \ac{UE} report information and \ac{UE} location assessment, so \ac{UE} privacy is not considered ~\cite{MTD}. Besides, channel estimation requires increased transmission bandwidth, which is a significant challenge in the current mobile communication standard. Therefore, the application of machine learning algorithms is appropriate in wireless channel modeling and estimation. The \ac{CH} area can be defined as any area in the environment and within a predefined transmitter coverage range, which does not meet the wireless channel reliability requirements~\cite{CH_iden}. The received signal power in the \ac{CH} is lower than the receiver's sensitivity, which leads to channel failure and increased packet error rate. The significant challenge for ensuring system stability in the real-time closed-loop controlled system is the data exchange between the sensors, actuators, and the edge cloud via highly reliable radio channels. Therefore, the \ac{CH} problem significantly impacts the stability of the networked controlled actuator. The trajectory follower is a class of closed-loop control systems that provides the ability to track a reference trajectory using error convergence between the current position and the reference trajectory, under the condition that the reliability of the transmission ensures the maintenance of closed-loop stability conditions~\cite{etciim22}. If the \ac{agv} could be located in an area of \ac{CH}, the connection outage can cause a delay in control input or the current sensor data update. Due to shadowing and multipath propagation, dynamic wireless coverage holes with high packet loss and packet delay probability are caused in the industrial environment. Therefore, increasing wireless signal propagation problems in the industrial environment requires a suitable solution to eliminate their effects of the \ac{agv} mobility.
\begin{figure*} [t]
\centering
\includegraphics[width=13cm,height=5.5cm]{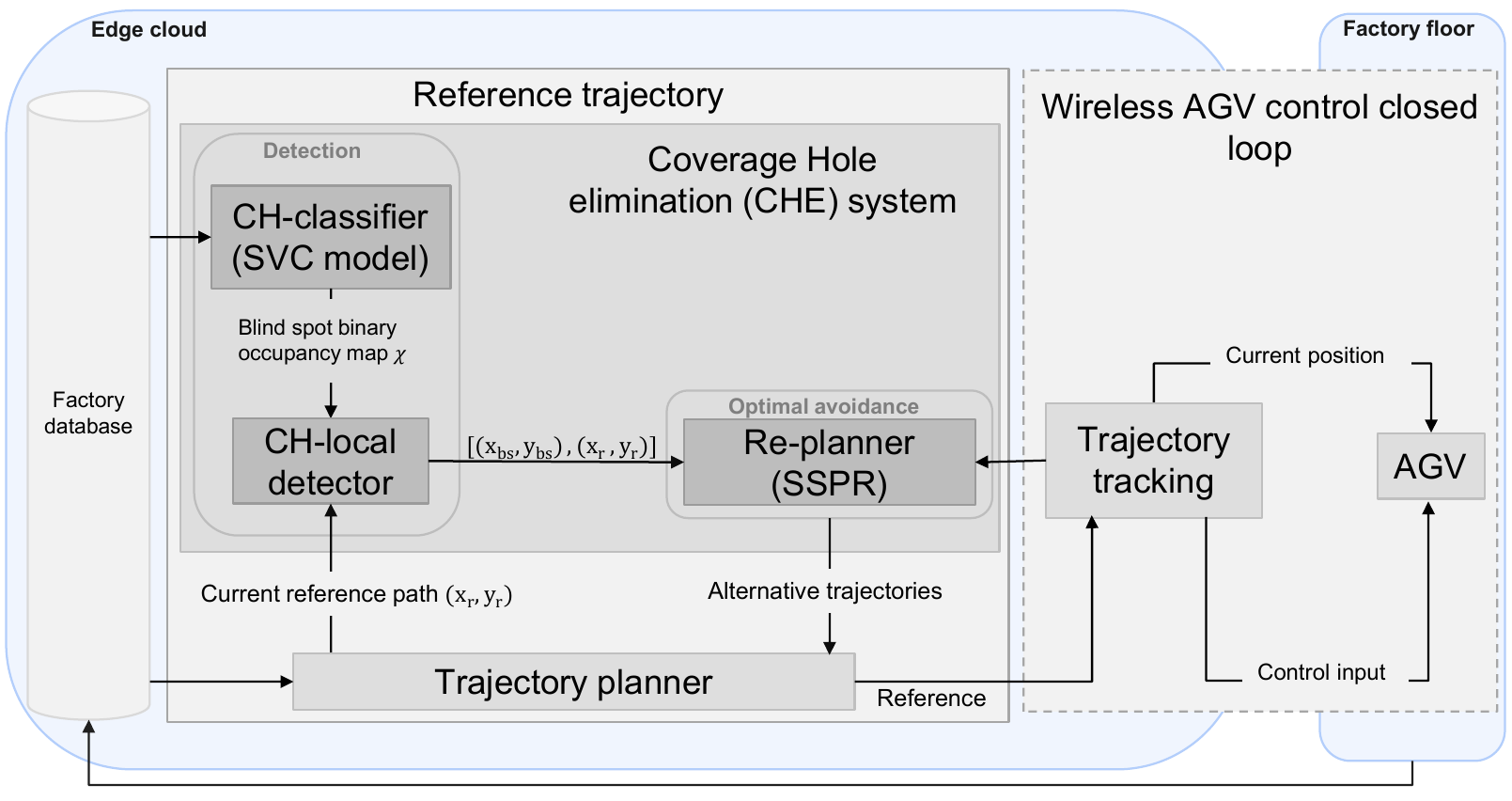} %
\caption{Enhanced edge cloud AGV trajectories follower with the \ac{CH} elimination (CBSE) system}   
\label{fig:1}
\end{figure*}
\subsection{Related work}
The occurrence of a  coverage hole in the current~\ac{agv} trajectory increases the number of successive packet drops to the tolerance limit~\cite{agv:stabil}, which increases the control error. If the AGV's deviation is more from the trajectory, it will not reach the target position and leads to the controller's instability. The author in~\cite{CHD_UAV} proposed a system to detect the  coverage hole by the drone without transmitting a user report via a backhaul link. The \ac{UAV} has two modes as a user and as a base station. The \ac{UAV}-as-user measures the received signal power and switches the \ac{UAV} to the second mode when a coverage hole is detected. The author in~\cite{GOCom} proposed a goal-oriented communication system based on the adaptive data rate aiming to control gain improvement. The author has modeled the packet delay and information freshness. Based on this, the wireless networked control system is optimized using an adaptive data rate. The author in~\cite{multi_connection_PER} proposes Multi-Link Connectivity (MLC) to support highly reliable, low-latency wireless links. The goal-oriented communication can be used as a solution for power map optimization. However, this conditions the knowledge of the signal strength map. The received signal strength map can be estimated via stochastic or deterministic indoor propagation model~\cite{radioMap}. The stochastic model has a low estimation accuracy since the exact indoor geometry is not considered. The deterministic method, like the ray tracing method, has a high accuracy level as the stochastic however, this method has relatively high computational time. The trajectory planning algorithms such as \ac{prm}, which use a sampled environment map, can effectively find the shortest trajectory from the start to the target position via search algorithms~\cite{prm_2022}. The detected coverage hole positions can be iteratively added to the sampled trajectory search map to find the optimal coverage-hole-free trajectory. In~\cite{re-plann}, the author proposed the avoidance concept with B\'ezier curve optimization for an unknown binary map environment via a trajectory re-direction around the blocked position using the close-set positions before and behind the blocked position. So in this work, the proposed trajectory re-plan algorithm finds the second optimal trajectory as an alternative trajectory, which describes the shortest coverage-hole-free trajectory to the target position.
\subsection{Contribution and organization of the paper}
The effect of coverage holes on \ac{agv} trajectory acquisition requires coverage hole position detection and an avoidance strategy. So this work proposes an ML-binary coverage hole map construction for the whole industry layout map iterative considering the layout or wireless transmission parameter change. So it is a binary representation of the detected coverage or non-coverage hole positions. Then, the coverage hole avoidance strategy proposes via an alternative trajectory re-planning algorithm. The trajectory re-planner maintains that the alternative trajectory is the second optimal trajectory in the sampled environment map. To the author's knowledge, such a coverage hole elimination system for edge cloud wireless trajectory control has yet to be proposed.

Section 2 provides an overview and a detailed description of the enhanced edge cloud AGV trajectories via the coverage hole elimination (CHE) system. Section 3 describes the implementation of the binary coverage hole map construction. Furthermore, the performances of coverage hole detection are analyzed. Section~4 presents the performance of trajectory tracking considering the re-planned trajectory. Finally, in Section~5, the paper is summarized.
\section{System Model} 
The \autoref{fig:1} describes the co-located \ac{cbse} system in the \ac{agv} wireless trajectory following. Each \ac{agv} is assisted via a wireless trajectory follower to track a predefined trajectory from a start position to a target position on the factory floor within a given time $T$. The data is exchanged  between the edge cloud and the industrial floor via wireless communication. The trajectory planner determines the optimal reference trajectory positions $(x_r,y_r)$, and the corresponding reference interpolated trajectories $T_r$ in the $(x,y)$-industrial coordinate system. A trajectory planner and follower are located in the edge cloud. Each AGV receives the necessary trajectory control information to follow the reference trajectory. The proposed \ac{cbse} system targets to detect the position of the \ac{CH} in the current \ac{agv} reference trajectory and to provide optimal alternative trajectories. The sample time interval of the \ac{cbse} system is $dt$. The coverage hole should be detected and avoided at each $k^{th}$ time step. The \ac{agv} obtains the control input from the trajectory follower at each $k^{th}$ time step. The industry layout is represented in the edge cloud as a binary map. The black pixels represent blocked positions, which cannot be entered by \ac{agv}, whereas the white pixels represent the free \ac{agv}-working area~\ref{fig:7a}. After coverage hole detection, the binary \ac{ch} map is constructed. Based on the \ac{ch}-local detector response, the current reference path is updated via the trajectory re-planner generating the alternative \ac{ch}-free trajectory as a reference input for trajectories follower.  
The reference trajectories $T_r$ consist of the reference trajectory positions $(x_r,y_r)$ in $(x,y)$-industry area coordinates, \ac{agv} orientation to the $x$-axis $(\theta_c)$ and the reference linear as well as angular velocity $(v_r,\omega_r)$. The \ac{ch} classifier constructs the \ac{ch} binary map. There are two classes of receiver power: The positions with a receiving power below the sensitivity threshold of the receiver are assigned to the \ac{CH}-class $(y=0)$. Otherwise, the position with sufficient receiving power above the receiver's sensitivity threshold is categorized as the non-\ac{CH}-class $(y=1)$. The industry map can be represented as a \ac{CH} binary map $\chi$ using each position's class assignment. So the position with the \ac{CH}-class is represented as a black pixel in the binary map. And the position with the non-\ac{CH}-class is described as a white pixel. As shown in \autoref{fig:1}, the \ac{CH} detection is performed in two steps. First, the \ac{CH} classifier should be able to assign the industry positions after training to two classes $\hat{y}\in \{0,1\}$. Then, a binary \ac{CH} map~$\chi$ can be created using industry position class assignments. Second, considering the current \ac{CH} binary map can determine the local detector if the current reference trajectory contains any \ac{CH}. Then, the position of the \ac{CH} $(x_{bs},y_{bs})$ should be submitted to the trajectories re-planner. That means the \ac{agv} has prior information about the \ac{CH} position, which can occur after a time interval \ac{lat} in reference positions. The trajectories re-planner creates alternative optimal reference trajectories considering the \ac{lat} between the current \ac{agv} position at the time step $k$ and the detected \ac{CH} position at the time step $k_{bs}$. Then updated reference trajectories are utilized at the trajectories follower to return a corresponding control input to \ac{agv} to minimize the error between the current \ac{agv} position and the reference position $(x_r,y_r,\theta_r)$. The input information for the \ac{CH} classifier is obtained from the industry database. This information should include the current industry conditions and the available RF-propagation information. Therefore, at each iteration $k^{th}$, the \ac{cbse} should detect the \ac{CH} in the current reference trajectory and then avoid them accordingly.
\section{Coverage hole detection}
 \begin{figure*} [t]
  \centering
\includegraphics[width=16cm,height=4cm]{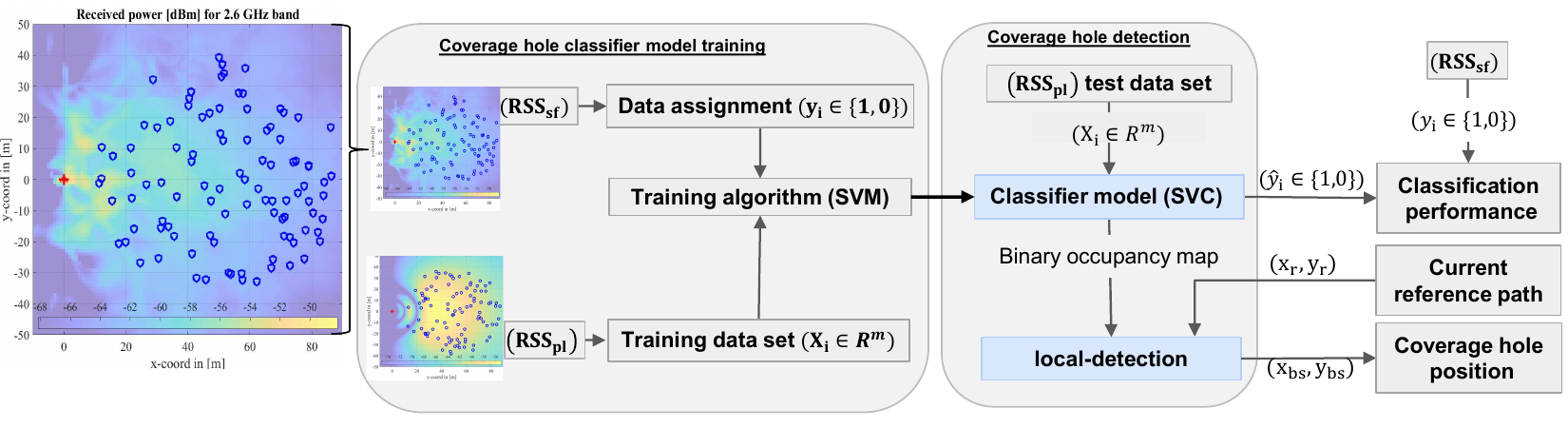}
  \caption{\ac{CH} classifier model (SVC), \ac{CH} binary map construction, \ac{CH} local-detection in the current reference path.}   \label{fig:3}
\end{figure*}
As shown in \autoref{fig:3}, the possible position of the \ac{CH} on the reference trajectory is determined in two steps. In the first step, a trained binary classifier  model should detect the \ac{CH} positions using support vector machine approach \ac{svm}. Each $i^{th}$ position has true class assignments ${y}\in \{0,1\}$ and detected class assignments from the classifier model $\hat{y}_i\in\{0,1\}$ in the training phase. Then, detected class assignments in the classifier test phase are considered to create the binary \ac{CH} map. The second step is local-determining the \ac{CH} in the current reference trajectory. The detected \ac{CH} coordinate position $(x_{bs},y_{bs})$ should be forwarded to the trajectories re-planner. The industry database has information for each position on the industry floor. This information is intended to keep the available information about the RF propagation at each time, the industry layout the operating frequency, the transmitter position, the transmitter height, and the mean RF path-loss at each position. However, the machine learning classifier will consider the RF signal loss by the multi-path and shadow fading as a binary \ac{CH} map. That means the trained classifier should find the correct position class assignment based on the data vector for each position. The \ac{svc} is trained with true position assignments ${y}\in\{0,1\}$ based on the \ac{rss} map in \ac{nlos} propagation model. Each $i^{th}$ position in the industry map will be described by a features vector $X_i$ in an m-dimensional feature space $R^m$. These features are provided as input information to the classifier for each position from the industry database. It should contain the RF transmission parameter and the available RF industry propagation conditions at each position using \ac{quadriga} channel model in the industrial environment, \cite{Quadriga:industry}. After that, the unlabeled data set is described as a matrix $[X_{ij}]_{n\times m};\  (i=1:n) $ observation positions, and $(j=1:m)$ with $m$ is the number of features. Each $i^{th}$ observation position should be assigned to one of the two classes $y_i=\{0,1\}$.
The class assignments $y$ are obtained from the received signal power map, including multi-path and shadow fading effects. Therefore, the training and validation phase determines the nonlinear hyperplane parameter. So the trained \ac{svc} can separate the positions of the industrial map in the m-dimensional features space $R^m$ using the labeled data set $[X_i,y_i]$. The trained \ac{svc} should then be able to provide the correct position assignment $\hat{y}_i\in\{0,1\}$ using the test data set. The  \ac{svc} response $\hat{y}_i$ describes binary position assignments that can be represented in the $(x,y)$ industry coordinate as a \ac{CH} binary map. In the second step, the current \ac{agv} reference trajectory $(x_r,y_r) \ \text{at} \ \text{each time step \ } (k=0:\frac{T}{dt}) $ is checked to determine if it has the potential to contain a \ac{CH}. As illustrated in \autoref{fig:3}, we have implemented the $\text{RSS}_{\text{pl}}$ map with consideration of the mean path-loss values using the empirical industrial channel model and the $\text{RSS}_{\text{sf}}$ map with consideration of the shadow and multi-path fading propagation effects. Both power maps in \autoref{fig:3} include a transmitter (red point) and multi-random distributed receivers (blue points) in an industry layout \cite{Quadriga:industry}. The input information $[X_{ij}]_{n\times m}$ is extracted from the $\text{RSS}_{\text{pl}}$ maps for the model training as well as for model testing. However, the position assigning in two categories $y_i\in\{0,1\}$ are obtained from the $\text{RSS}_{\text{sf}}$ map. The labeled test data set is also necessary to analyze the model performance by comparing the correct class assignment $y_i$ from the $\text{RSS}_{\text{sf}}$ map and the class assignments form the classifier $\hat{y}_i$. The separation of the two classes requires a nonlinear hyperplane feature space $R^m$ using \ac{rbk} kernel. The penalty parameter in the \ac{svc} represents the inverse effect of the regularization parameter. The penalty parameter and kernel parameter are determined by L-fold cross-validation.
\subsection{Channel measurement}
For a specific industry layout, the received signal power can be changed depending on several parameters, e.g., the variation of the height and the position of the transmitter, as well as the operating frequency and the distance to the transmitter. The mean path-loss value under change of these parameter changes should obtain at the model training from the path-loss map $\text{RSS}_{\text{pl}}$ map where $\text{RSS}_{\text{pl}}$ map is obtained considering Hata's model in the \ac{quadriga} industrial scenario. Furthermore, the data assignment is done from the $\text{RSS}_{\text{sf}}$ industrial map considering the shadowing effect as well as the multi-path effect. A random decrease in the signal strength occurs due to reflections, scattering from metallic surfaces, and multiple duplicates of the transmitted signal arrive. At the receiver, multiple paths lead to  variations in the received signal strength. consequently a \ac{CH} can occur in the $\text{RSS}_{\text{sf}}$ coverage map. As shown in \autoref{fig:3}, the area around $(25,-20)$ is the industry $\text{RSS}_{\text{sf}}$ map with height loss of received signal power. Although, in the $\text{RSS}_{\text{pl}}$ map, this wireless channel loss does not appear in this area. The training data set is assigned with the correct position assignments via the $\text{RSS}{_{\text{P}_{\text{sf}}}}$ value. After that, the trained \ac{svc} should be able to assign \ac{CH} positions as a positive class based on the available input information from the industry floor. As in \autoref{fig:3} show, the \ac{quadriga},\cite{Quadriga:industry}, is used to create the \ac{rss} map; the red position is the transmitter with multiple randomly distributed receivers.
\subsection{Coverage hole detection performance}
\begin{figure}[t]
\centering
\includegraphics[width=6cm,height=4.5cm]{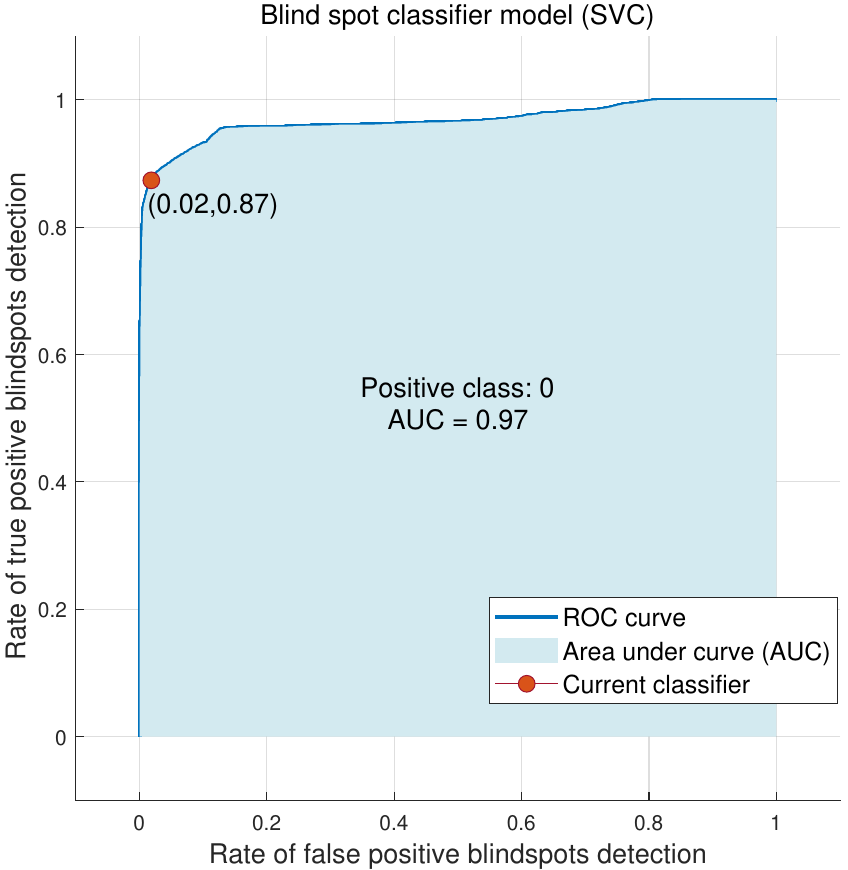}
  \caption{Receiver operating characteristic (ROC) of the trained model}
  \label{fig:5}
    \end{figure}

\begin{figure}[t]
\centering
\includegraphics[width=0.7\linewidth]{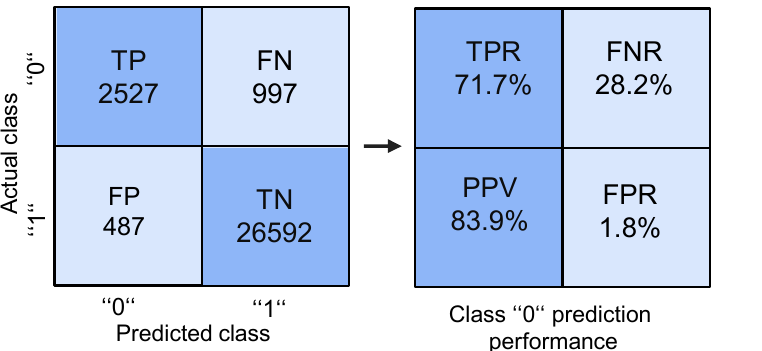}
  \caption{Confusion matrix of the model test and \ac{CH} detection performance; TP: true positive, FN: false negative.}
  \label{fig:6}
\end{figure}

Since the \ac{CH} classifier is used to separate imbalanced binary classes based on the available database information. The \ac{roc} curve is proposed to determine the operating point for the classifier model as shown in \autoref{fig:5}. The true positive and the false positive \ac{CH} class ''0'' prediction rate is calculated with a variable prediction threshold. However, with the optimal threshold selection, the overfitting problem is avoided. On the other side, the trade-off between true and false decisions, the \ac{CH}, is established. The free-\ac{CH} area can be reduced if the \ac{CH} false alarm detection is higher. The classifier has 0.97 \ac{auc}, which means that our classifier has the potential to predict the \ac{CH} class correctly but still needs to find the optimal operation point which balances the false and true positive rate. The geometric mean between the false positive rate (FPR) and the true positive rate (TPR) is used to detect it. The operation point of the classifier is determined with 87$\%$ true positive rate as shown in\autoref{fig:5} as a red point. Nevertheless, it is accompanied by a 2$\%$ false positive \ac{CH} prediction rate. However, as shown in \autoref{fig:6}, the classifier performance is analyzed regarding the \ac{CH} class so that the \ac{CH} detection metrics are extracted from the classification confusion matrix. As expected, the classifier model can predict the \ac{CH} class with a true positive rate $\text{TPR}=\frac{TP}{TP+FN}=71.7\%$. However, this is also accompanied by the false positive \ac{CH} prediction rate $\text{FPR}=\frac{FP}{TN+FP}=1.8\%$.
The increase in the accuracy of \ac{CH} prediction is achieved with an increased prediction rate of false alarm, which reduces the working space of \ac{agv}. Nevertheless, this trade-off between TPR and FPR is limited within the \ac{roc} curve.
\section{Coverage hole avoidance}
\begin{figure*} [t]
\centering
          \subfloat[ ][Industry \ac{CH} binary map\cite{etciim22}.]{\includegraphics[width=0.4\textwidth,height=6cm]{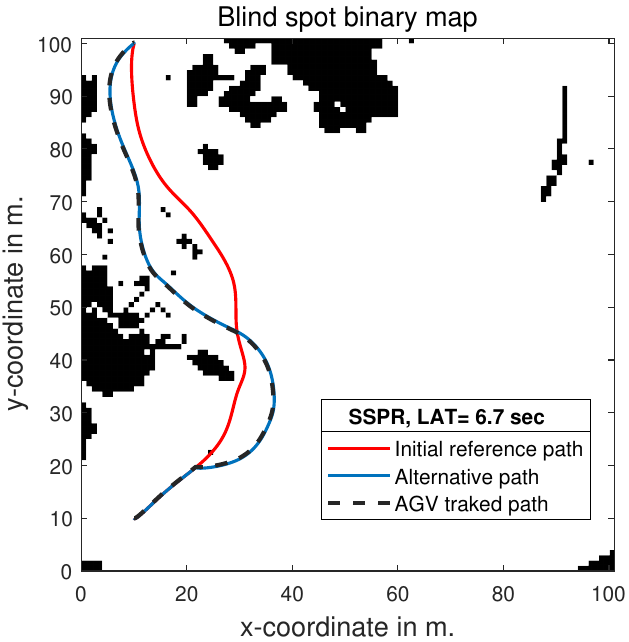}
         \label{fig:7a}} \hspace{0.9cm}
         \quad
          \subfloat[][The trajectory follower error.]{\includegraphics[width=0.4\textwidth,height=6.35cm]{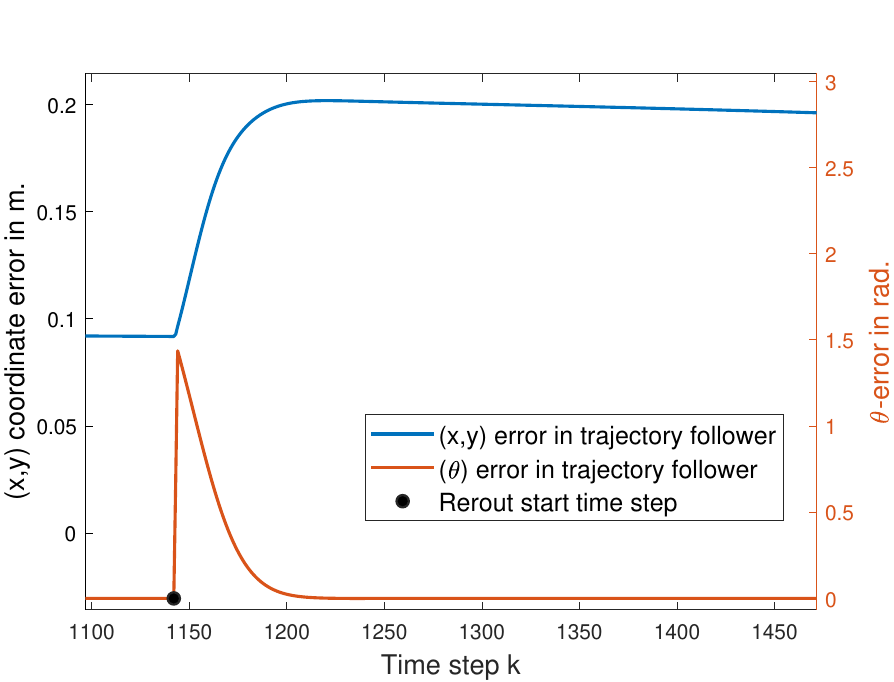}
         \label{fig:7b}}
        \caption{Trajectories planning, re-planning, and following.}
        \label{fig:7}
\end{figure*} 
When a \ac{CH} occurs in the current $(x,y)$-reference trajectory, the trajectories re-planner should find alternative optimal reference trajectories based on the updated \ac{CH} binary map from the \ac{CH} classifier. The trajectory re-planner should find the optimal trajectory re-routing start $SP_{re}$ to the initial target position, considering the $\text{LAT}$ between the current \ac{agv} position and the \ac{CH} position. Then the alternative trajectory positions should interpolate as $(x_r,y_r)$ and from it should calculate the appropriate reference \ac{agv} orientation $\theta_r$, the linear as well as the winkle velocity $(v_r,w_r)$.

\subsection{Trajectory planner using PRM}
 Using \ac{prm} the optimal initial reference trajectory is generated. A search algorithm \ac{astar} can find an optimal trajectory with minimal length in the PRM map under given start and target positions. The implementations of \ac{prm} are categorized into two phases a preparation phase and a query phase. In the preparation phase, a PRM map is constructed using generated random nodes $N$ in the industry map's free \ac{agv} workspace and the connection from each node to its nearest neighbors by a straight line distance smaller than the radius $D_{max}$. The possible edges $E$ between nodes $N$ build the PRM map. Then a graph search is performed using \ac{astar} in PRM map to find optimal trajectory points.
Therefore, the reference trajectories are all predefined trajectory information as reference data extracted from $(x,y)$-trajectory points using the trajectory interpolation optimization. The linear and angular velocity should be customized under the technical conditions of the \ac{agv}. 
The cubic \ac{B-spline} interpolation with B\'ezier function $\Psi(\cdot)$ can interpolate the variable length trajectory segments  \cite{bspline}. So, each segment is optimized with the B\'ezier curve.
\subsection{Trajectories re-planner using SSPR}
The optimization of the alternative trajectory is accomplished considering the look-ahead time. The look-ahead time defines the time range in which the system should find an optimal alternative. The algorithm finds the optimal alternative trajectory in the corresponding PRM map based on the current \ac{CH} binary map $\chi$. Then, using the proposed \ac{ssr} algorithms a start position for the trajectory re-route $SP_{re}(t=t_{re})$ should be determined to provide a shorter alternative trajectory from the initial start position to the target position. As illustrated in \autoref{fig:7a}, \ac{CH} positions are classified at the time step $k$, and the \ac{CH} binary map is updated. A start position for trajectory re-routing is planned within the look-ahead time. The alternative optimization method guarantees that the \ac{agv} follows the first optimal trajectory within a specific time before starting the avoidance process. 
\autoref{fig:7a} presented a \ac{CH} binary map $\chi(k)$ as an example of the classifier output at the time step $k$. The initial trajectory, which is presented in red, is planned in the previous coverage hole binary map $\chi(k-1)$; thereby, the position at $(20,21)$ was assigned to the \ac{ch}-free workspace. Then, the \ac{ch} map is updated at the next time step $(k)$, and a coverage hole is determined at $(20,21)$. The \ac{CH} detector gives the \ac{CH} position at  time step $k_{bs}$ to the trajectories re-planner..
Our proposed \ac{ssr} algorithm considers the possible start positions $(SP_{re}=(x_{re},y_{re})$ for the trajectory re-routing within the look-ahead time $LAT =(dt(k-k_{bs}))$ to minimize the total \ac{agv} alternative trajectory. That is shown in \autoref{fig:7a}, that \ac{agv} tracks the initial trajectory and then before the \ac{CH} in $1 \si{\sec}$ starts with the trajectory re-route. This alternative trajectory length is minimized within LAT=$6.7 \si{\sec}$ using \ac{ssr}. The trajectory follower should minimize at each time step the coordinate error between the reference trajectory positions (the alternative in blue) and the tracked \ac{agv} trajectory positions (in black). The trajectory following error in the (x,y)-coordinates and the \ac{agv} orientation is shown in \autoref{fig:7b}. The trajectory follower can converge the orientation error after $0.3  \si{\sec}$. However, the (x,y)-coordinate error increases to the target by $0.11  \si{\meter}$. The \ac{agv} can avoid the \ac{CH} and reach the target position. However, this may increase the task time $T$ if the additional distance to the target cannot be corrected with the increase of the \ac{agv} linear velocity $v$. Furthermore, the re-route increases the trajectory following (x,y)-error in the centimeter range. 
\subsection{Trajectory re-planning performance}
\begin{figure}[t]
\centering
    \includegraphics[width=8.5cm,height=6.5cm]{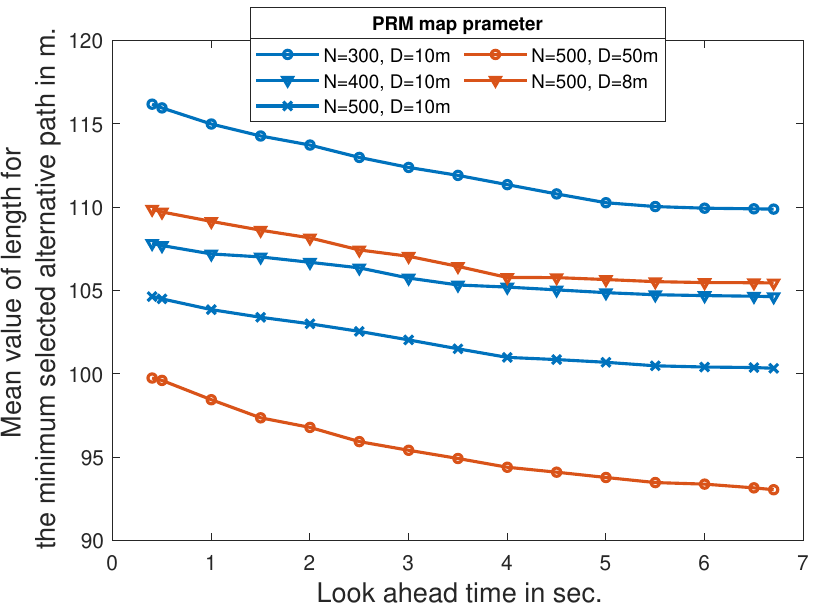}
  \caption{The alternative trajectory length with various LAT.}
  \label{fig:10}
    \end{figure}

This section analyzes the performance of the \ac{ssr}. The \ac{agv} should avoid the \ac{CH} position and reach the target position with an acceptable trajectory coordinate error. Under the condition of the \ac{agv} speed limitation $v_{max}$, the  re-planning performance depends on the \ac{lat} is analyzed, as can be seen in \autoref{fig:10}. This is expected that greater \ac{lat} provides more trajectory re-routing possibilities. The optimal alternative trajectory strongly depends on \ac{lat} and is getting reduced with increased \ac{lat}; if we make a \ac{CH} detection earlier, we gain more possibilities to achieve more optimization at \ac{CH} avoidance. Conversely, The \ac{lat} depends on how often the \ac{CH} on the \ac{agv} trajectory occurs and disappears. In \autoref{fig:10}, the analysis of multiple \ac{lat} is executed at $10$ simulations, and the average trajectory length is plotted. The randomly distributed node position is changed. Thus the PRM map is updated at each experiment due to the change of node positions. The target position remains the same in all experiments. It is clear that at larger \ac{lat}, the additional distance due to the trajectory re-route is reduced, and the alternative trajectory is minimized. Furthermore, in \autoref{fig:10} is shown that with the increase of the road map parameter $N$ and $D_{max}$ the alternative trajectory length is minimized for the same remaining \ac{lat}. Nevertheless, the increase of road map parameters is limited because of the computation resources and the real-time system requirements. Consequently, the earlier the coverage hole is identified, the shorter the alternative trajectory can be found. But if the coverage hole position is assigned to the free \ac{agv} workspace after the re-route trajectory to start. Then, the \ac{lat} should decrease since the multiple re-routing increases the trajectories following coordinate error. However, the multiple re-routing can be optimized by adjusting the PRM map parameters and the probability of coverage hole occurrence and disappearance. For highly dynamic RF propagation changes, the map parameters should increase to avoid the coverage hole considering the reduction of the alternative trajectory length and the tracked trajectory coordinates error.
\section{Conclusion}
The proposed \ac{cbse} system can detect and avoid the coverage hole. We evaluate the performance of coverage hole elimination in wireless controlled \ac{agv} use case in an industry environment. With environment layout dependence, the trained \ac{svc} model can identify the coverage hole position with high detection performance. Then, the detected coverage hole is successfully avoided via the suggested SSPR trajectory re-planner algorithm. The \ac{agv} follows the optimal alternative trajectory and reaches the target position. The larger LAT can improve coverage hole avoidance performance considering the \ac{agv} trajectory length. Moreover, the re-planning performance regarding the alternative trajectory length can also be enhanced by increasing the PRM map parameters. However, the PRM map parameter increase is constrained due to the computational edge cloud resources and real-time conditions. Nevertheless, the \ac{lat} should adjust with the coverage hole change frequency because the unnecessary avoidance with larger \ac{lat} can increase the AGV coordinate error.

\section*{\uppercase{Acknowledgements}}
The authors acknowledge the financial support by the Federal Ministry of Education and Research of the Federal Republic of Germany (BMBF) in the project “Open6GHub” with funding number (DFKI-16KISK003K) and in the project “AIRPoRT” with funding number (01MT19006A). The authors alone are
responsible for the content of the paper.

\bibliographystyle{IEEEtran}
\bibliography{coverageHole.bib}
 
\end{document}

%% file: IEEEtran/acroList.tex
\begin{acronym}[MPC] 
\acro {agv}[AGV]{Automated Guided Vehicle} 
\acroplural {agv}[AGVs]{Automated Guided Vehicles} 
\acro {svc}[SVC]{Support Vector Classifier}
\acro {qos}[QoS]{Quality of Service}
\acro {rem}[REM]{Radio Environment Map}
\acroplural {rem}[REMs]{Radio Environment Maps}
\acro {dnn}[DNN]{Deep Neural Network}
\acro {dl} [DL] {Deep Learning}
\acro {ris} [RIS]{Reconfigurable Intelligent Surface}
\acroplural {ris} [RISs]{Reconfigurable Intelligent Surfaces}
\acro {crs}[CRS]{Cognitive Radio System}
\acro {mse}[MSE]{Mean Squared Error}
\acro{ann}[ANN] {Artificial neural networks}
\acro {pca}[PCA]{Principal Component Analyse}
\acro{mlp-ann}[MLP-ANN]{multi layer perception-Artificial neural networks}
\acro{svr}[SVR]{Support Vector Regression}
\acro{svm}[SVM]{Support-Vector Machine}
\acro{sl}[SL]{supervised learning}
\acro{k-nn}[k-NN]{k-Nearest Neighbours}
\acro{dt}[DT]{Decision Trees}
\acroplural{svm}[SVMs]{Support-Vector Machines}
\acro {quadriga}[QuaDRiGa] {QUAsi Deterministic RadIo channel GenerAtor}
\acro{los} [LOS] {Line-of-Sight}
\acro {nlos} [NLOS]{Non-Line-of-Sight}
\acro {rbk}[RBK]{Radial Basis Kernel}
\acro {ssr}[SSPR] {Start Selection Path Re-planning}
\acro{prm}[PRM]{Probabilistic Road-maps}
\acro{lat}[LAT]{Look-Ahead Time}
\acro {astar}[A*]{A-Star}
\acro{roc}[ROC]{Receiver Operating Characteristic}
\acro{fdr}[FDR]{False Discovery Rate}
\acro {fnr}[FNR]{False Negative Rate}
\acro {tpr}[TPR]{True Positive Rate}
\acro{auc}[AUC]{Area Under Curve}
\acro {B-spline} [B-spline] {Basis Spline}

\acro {fp}[FP]{False Positive}
\acro {tn}[TN]{True Negative}
\acro {fn}[FN]{False Negative}
\acro {ppv}[PPV]{Positive Predictive Value} 
\acro {fnr}[FNR] {False Negative Rate}
\acro {tpr}[TPR] {True Positive Rate}
\acro{fpr}[FPR]{False Positive Rate }
\acro {tp}[TP]{True Positive}
\acro{cbse} [CHE]{Coverage Hole Elimination}
\acro{ch} [CH]{Coverage Hole}
\acro{rss} [RSS]{Received Signal Strength}
\acro{SP}[sp] {Start Position}
 \acro{UAV}[UAV] {Unmanned aerial vehicle}
 \acro{UE}[UE] {User Equipment} 
 \acro{CH}[CH] {Coverage Hole} 
\end{acronym}